\documentclass[12pt,a4paper,twoside]{article}

\usepackage[dvips]{graphics}
\usepackage{cite}

\newcommand{\qqbar}  {\ensuremath{\mathrm{q\overline{q}}}}
\newcommand{\uubar}  {\ensuremath{\mathrm{u\overline{u}}}}
\newcommand{\ddbar}  {\ensuremath{\mathrm{d\overline{d}}}}
\newcommand{\ssbar}  {\ensuremath{\mathrm{s\overline{s}}}}
\newcommand{\bbbar}  {\ensuremath{\mathrm{b\overline{b}}}}
\newcommand{\ccbar}  {\ensuremath{\mathrm{c\overline{c}}}}

\newcommand{\epem}   {\ensuremath{\mathrm{e^+e^-}}}

\newcommand{\as}     {\ensuremath{\alpha_s}}
\newcommand{\asu}     {\ensuremath{\alpha_s^\mathrm{uds}}}
\newcommand{\asc}     {\ensuremath{\alpha_s^\mathrm{c}}}
\newcommand{\asb}     {\ensuremath{\alpha_s^\mathrm{b}}}
\newcommand{\rasbasu} {\ensuremath{\alpha_s^\mathrm{b}/\alpha_s^\mathrm{uds}}}
\newcommand{\rascasu} {\ensuremath{\alpha_s^\mathrm{c}/\alpha_s^\mathrm{uds}}}
\newcommand{\mb}     {\ensuremath{\mathrm{M}_\mathrm{b}}}

\newcommand{\oa}     {\ensuremath{\mathcal{O}(\alpha_s)}}
\newcommand{\oaa}    {\ensuremath{\mathcal{O}(\alpha_s^2)}}

\newcommand{\znull}  {\ensuremath{\mathrm{Z^0}}}

\newcommand{\xmu}    {\ensuremath{x_{\mu}}}

\newcommand {\downto}
         {\mbox{ \begin{picture}(14,10)
                    \put(0,10){\line(0,-1){5.0}}
                    \put(2,5){\oval(4,4)[bl]}
                    \put(1,0){\makebox(0,0)[bl]{$\rightarrow$}}
                 \end{picture} }}

\newcounter{hours}
\newcounter{minutes}

\newcommand{\Printtime}{%
  \setcounter{hours}{\time/60}%
  \setcounter{minutes}{\time-\value{hours}*60}%
  \ifthenelse{\value{hours}<10}{0}{}\thehours:%
  \ifthenelse{\value{minutes}<10}{0}{}\theminutes}

\begin{document}

%
\begin{titlepage}
\begin{center}{\large   EUROPEAN LABORATORY FOR PARTICLE PHYSICS
}\end{center}\bigskip
\begin{flushright}
       CERN-EP/99-045 \\     
March 19, 1999
\end{flushright}
\bigskip\bigskip\bigskip\bigskip
\begin{center}
{\Huge\bf
Test of the Flavour Independence of {\bf $\alpha_s$}  using
Next-to-Leading Order Calculations for Heavy Quarks
}
\end{center}
\bigskip\bigskip
\begin{center}
{\large The OPAL Collaboration}
\end{center}
\bigskip

%
\begin{abstract}
  We present a test of the flavour independence of the strong
coupling constant for charm and bottom
quarks with respect to light (uds) quarks, based on a hadronic 
event sample obtained with the OPAL detector at LEP. 
Five observables related to global event shapes 
were used to measure \as\ in three flavour
tagged samples (uds, c and b). 
The event shape distributions were fitted by \oaa\ calculations
of jet production taking into account mass effects for the c and b quarks.
We find: 
\begin{center}
~~~~~$\rascasu~= 0.997 
  \pm 0.038  ~(stat.) \pm 0.030 ~(syst.) \pm 0.012 ~(theory)$ and 
$\rasbasu~= 0.993 
  \pm 0.008  ~(stat.) \pm 0.006 ~(syst.) \pm 0.011 ~(theory)$. 
\end{center}
\end{abstract}

\bigskip\bigskip\bigskip
\begin{center}{\large
(To be submitted to Eur. Phys. Jour. C)
}\end{center}
 
\end{titlepage}

\begin{center}{
G.\thinspace Abbiendi$^{  2}$,
K.\thinspace Ackerstaff$^{  8}$,
G.\thinspace Alexander$^{ 23}$,
J.\thinspace Allison$^{ 16}$,
N.\thinspace Altekamp$^{  5}$,
K.J.\thinspace Anderson$^{  9}$,
S.\thinspace Anderson$^{ 12}$,
S.\thinspace Arcelli$^{ 17}$,
S.\thinspace Asai$^{ 24}$,
S.F.\thinspace Ashby$^{  1}$
% ...
}\end{center}\bigskip
\bigskip
$^{  1}$School of Physics and Astronomy, University of Birmingham,
Birmingham B15 2TT, UK
\newline
$^{  2}$Dipartimento di Fisica dell' Universit\`a di Bologna and INFN,
I-40126 Bologna, Italy
\newline
% ...
$^{  5}$Cavendish Laboratory, Cambridge CB3 0HE, UK
\newline
% ...
$^{  8}$CERN, European Organisation for Particle Physics,
CH-1211 Geneva 23, Switzerland
\newline
$^{  9}$Enrico Fermi Institute and Department of Physics,
University of Chicago, Chicago IL 60637, USA
\newline
% ...
$^{ 12}$Indiana University, Department of Physics,
Swain Hall West 117, Bloomington IN 47405, USA
\newline
% ...
$^{ 16}$Department of Physics, Schuster Laboratory, The University,
Manchester M13 9PL, UK
\newline
$^{ 17}$Department of Physics, University of Maryland,
College Park, MD 20742, USA
\newline
% ...
$^{ 23}$Department of Physics and Astronomy, Tel Aviv University,
Tel Aviv 69978, Israel
\newline
$^{ 24}$International Centre for Elementary Particle Physics and
Department of Physics, University of Tokyo, Tokyo 113-0033, and
Kobe University, Kobe 657-8501, Japan
\newline
% ...
\newline
%
%
\newpage
\section{Introduction}

According to the theory of strong interactions, Quantum 
Chromodynamics (QCD)~\cite{qcd}, the strong coupling constant \as\ 
is the same for all quark flavours. Therefore a precise
measurement of \as\ for the individual quark flavours is an 
important test of this theory. The coupling constant \as\ 
for charm and bottom
quarks can be compared to \as\ for light (uds) quarks
by measuring the ratios \rascasu\ and \rasbasu\ using 
hadronic events of the type $\epem \rightarrow \qqbar g$. 
The coupling constants are measured   
using an event sample originating from a pair of light
quarks (\uubar , \ssbar\ or \ddbar),
c quarks (\ccbar) or b quarks (\bbbar), respectively.
If either of the above  ratios
deviates significantly from unity then this may indicate
physics beyond the Standard Model. 

In many QCD studies at LEP corrections due to quark mass effects 
can be safely ignored
because they typically appear as powers of the ratio
of the quark mass to the total energy. However, in those studies
where \as\ is determined in event samples enriched in heavy
quarks, mass effects become non-negligible. 
Gluon emission by bottom quarks, and
to a lesser extent charm quarks, will be suppressed largely due
to the reduced phase-space available.         
Observables sensitive to the three-jet rate
measured in heavy quark events will be modified with respect
to the same quantities measured in light quark events.

Tests of the flavour independence of \as\ have previously been
conducted at both LEP and SLC 
~\cite{flvtstopal1,flvtstopal2,flvtstl3,flvtstdelphi,flvtstaleph,flvtstsld} 
using ratios of \as\ for one flavour over \as\ of either the 
complementary
\footnote{For example, \as\ for c quarks was 
          compared to \as\ for a mixture
          of u, d, s and b quarks.} 
or inclusive
quark mixture. In all these tests flavour independence of \as\ 
was confirmed. These tests, however, used either massless QCD
calculations or a leading order calculation with massive quarks~\cite{heavyQ0}.
The leading order calculation includes
the process $\epem \rightarrow \qqbar gg$, 
with mass effects,  but virtual
corrections to the process $\epem \rightarrow \qqbar g$
are not included.
 
Complete next-to-leading-order (NLO) calculations
of the heavy-flavour production cross section in \epem\ collisions,
including quark mass effects,
have been published recently~\cite{heavyQ1, heavyQ2, heavyQ3}. 
A comparison between these calculations
has been made and they were found to be in agreement~\cite{heavyQ3}. 
In recent publications DELPHI~\cite{delphi_massive} and
SLD~\cite{sld_massive} report
on measurements of \rasbasu\ where NLO 
massive calculations~\cite{heavyQ1,heavyQ2} were used to account for 
mass effects in b quark events. These results are consistent 
with the flavour independence of \as .

In this study we use the  
results of P.~Nason and C.~Oleari~\cite{heavyQ3}
along with theoretical
predictions assuming massless quarks in fits to 
global event shape distributions
in order to determine \asu\ , \rascasu\ and \rasbasu. 
The results presented here are intended to update and
supersede the corresponding OPAL results in~\cite{flvtstopal1,flvtstopal2}
insofar as we now have greatly increased the charm event statistics and 
have used improved theoretical predictions.

  This paper is organised as follows. In Sect.~\ref{sec:dete}
the parts of the OPAL detector most important to this analysis
are described. In Sect.~\ref{sec:evt} the hadronic event sample
and the Monte Carlo event sample are introduced.
In Sect.~\ref{sec:buds} and \ref{sec:dstar} the
flavour tagging methods are
described. In Sect.~\ref{sec:shp} the event shape 
observables used in this study are
introduced and the procedure for correcting the
event shape distributions is explained.
Next, in Sect.~\ref{sec:oas2}, the procedure for fitting the 
NLO QCD prediction to the corrected distributions is explained and
in Sect.~\ref{sec:sys} the systematic uncertainties that
have been taken into account are
discussed.  In Sect.~\ref{sec:res},
the results of the test of flavour independence of \as\ are
presented.           
Finally, in Sect.~\ref{sec:con}, conclusions are drawn.

\section{The OPAL detector}
\label{sec:dete}

  The OPAL detector operates at the LEP \epem\ collider at CERN. A
detailed description can be found in
Refs.~\cite{opaldete,si}. The analysis presented here
relies mainly on the reconstruction of charged particle trajectories
and momenta in the
central tracking chambers, on energy deposits (``clusters'') in 
the electromagnetic calorimeters and on information from
the silicon micro-vertex detector.
 
All tracking systems are located inside a solenoidal magnet which
provides a uniform magnetic field of 0.435~T along the beam
axis\footnote{In the OPAL coordinate system the $x$ axis points
  towards the centre of the LEP ring, the $y$ axis points upwards and
  the $z$ axis points in the direction of the electron beam.  The
  polar angle $\theta$ and the azimuthal angle $\phi$ are defined
  with respect to the  $z$- and $x$-axes, respectively, while $r$ 
  is the distance from the
  $z$-axis.}.
The magnet is surrounded by a lead glass electromagnetic
calorimeter and a hadron calorimeter of the sampling type. Outside
the hadron calorimeter, the detector is surrounded by a system of 
muon chambers. There are similar layers of detectors in the barrel
($|\rm{cos}\theta | < 0.82$) and endcap ($|\rm{cos}\theta | > 0.81$)
regions.
 
The central tracking detector consists of a silicon micro-vertex 
detector~\cite{si} and three drift chamber devices: the vertex 
detector, 
a large jet chamber, and  surrounding $z$-chambers. The silicon 
micro-vertex detector, close to the beam pipe, consists of two 
layers of silicon strips with a single-hit resolution of about 
7\,$\mu$m in the $r\phi$ plane.
The vertex chamber is a cylindrical drift chamber covering
a range of $|\cos\theta|<0.95$. Its single hit resolution is 
50\,$\mu$m in the $r\phi$ plane and 700\,$\mu$m in the $z$ direction.
The jet chamber is a cylindrical drift chamber with an inner radius 
of 25\,cm, an outer radius
of 185\,cm, and a length of about 4\,m. 
Its spatial resolution is about 135\,$\mu$m in the $r\phi$ plane 
from drift time information and about 6\,cm in the $z$ direction 
from charge division. The $z$-chambers
provide a more accurate $z$ 
measurement with a resolution of about 300\,$\mu$m. In combination,
the three drift chambers yield a momentum resolution of
$\sigma_{p_t}/p_t \approx \sqrt{0.02^2+(0.0015\cdot p_t)^2}$ 
for $|\cos(\theta)| < 0.7$, where $p_t$ is the transverse momentum 
in GeV/$c$. 

Electromagnetic energy is measured by lead glass calorimeters
surrounding the solenoid magnet coil. They consist of a barrel and 
two endcap arrays 
with a total of 11704 lead glass blocks covering a range of 
$|\cos\theta|<0.98$. 

\section{Event sample and Monte Carlo simulation}
\label{sec:evt}

This analysis is based on a sample of 4.4 million hadronic 
decays of the \znull\ recorded with the OPAL
detector between 1990 and 1995.             
Hadronic \znull\ decays were selected by placing requirements on
the number of reconstructed tracks and the energy deposited
in the calorimeter. A detailed description of the criteria
is given in~\cite{bib-OPALmh}.
The parts of the detector essential for the present analysis
(central detector and electromagnetic calorimeter) were required to be
fully operational.  
The track selection criteria were the same as presented in a
previous OPAL study~\cite{opalresummed}.
The number of accepted
tracks was required to be at least five to reduce $\tau^+\tau^-$ 
background. Clusters of electromagnetic
energy were used if their observed energy was greater than 0.25\,GeV, 
and known noisy channels in the detector were ignored.
The event thrust axis~\cite{opalresummed} was determined using
all accepted tracks and clusters, and its direction was 
required to fulfil the condition $|\cos\theta_\mathrm{Th}|<0.9$ in
order that the event be well contained.  Using these
selection criteria, Monte Carlo studies indicate that, within the
chosen range of $\cos\theta_\mathrm{Th}$,
99.86$\pm$0.07\% of hadronic \znull\ decays are accepted, with a 
contamination
of about 0.14\% from $\tau^+\tau^-$ events, and around 0.07\% from
two-photon interactions~\cite{opalresummed}.

  To correct the measured event shape distribitions (see 
Sect.~\ref{sec:cor}), 4 million
hadronic decays of the \znull\ have been simulated using the
JETSET 7.4 Monte Carlo model~\cite{bib-JETSET}
with parameters tuned to represent OPAL data well
\cite{bib-OPALtune}. For all simulated events heavy quark fragmentation
has been implemented using the model of 
Peterson et al.~\cite{bib-PETERSON}.
All events have been  passed through a detailed
simulation of the OPAL detector~\cite{bib-OPALGOPAL}
before being analysed using the same programs as for data.

\section{Selection of uds and b quark events}
\label{sec:buds}

  Two event samples, one enriched in uds quark events and one      
enriched in b quark events, were selected by first determining the number 
of tracks in each event ($N_{sig}$) with a large impact parameter
significance, $b/\sigma_b > 2.5$. Here $b$ is the distance of closest approach in 
the $xy$ plane of the track 
to the \epem\ interaction point (IP) and $\sigma_b$ its error. 
The sign of $b$ was determined with respect to the crossing
point between the track and the jet axis.   
$b$ is positive if the track crosses the jet axis 
downstream of the IP and negative otherwise.             
An event was classified as having a high probability
to have come from light quarks
if $N_{sig}$ was zero and from b quarks if
$N_{sig} \geq 5$. 
In the text below these will be referred to as the ``uds-tag'' and 
the ``b-tag'' event samples.

  For the uds and b quark event selection all data recorded
during 1994 were used. This represents 1.4 million events. 
The analysis was restricted to only the 1994 data because of the 
uniform configuration of the silicon micro-vertex detector 
during this time period.
In addition to the track selection criteria outlined above,
the track was
required to contain at least one silicon hit and 
an algorithm was applied to reject tracks which were
consistent with arising from photon conversions~\cite{bib-idgcon}.
In order for the event to be contained within the acceptance
of the silicon micro-vertex detector
the event's thrust axis was
further restricted to lie within the range
$|\cos\theta_\mathrm{Th}| < 0.7$.      

  The distribution of $N_{sig}$ is shown in Fig.~\ref{fig:nsig} 
compared with the result of Monte Carlo simulation. 
The simulation is decomposed into the contributions from uds, c 
and b quark events. The general agreement between the sum of the
three contributions from Monte Carlo and the data is good. 

  This tagging procedure resulted in 325\,111 events selected for
the uds-tag event sample and 71\,521 for the b-tag event sample.
The flavour compositions of the uds-tag and the b-tag event samples
as determined from the Monte Carlo, along with their combined 
statistical and systematic errors,
are presented in Table~\ref{tab:flavor}.
The efficiency for tagging uds events was about
35\% and for tagging b events 23\%.

\subsection{Uncertainties in Flavour Composition}

  The systematic errors on the flavour compositions result from
uncertainties in the detector modelling and imprecise knowledge
of physics processes. For each source of uncertainty some
aspect of the Monte Carlo was varied and the flavour
composition of the uds-tag and b-tag event samples were 
recalculated. The difference between the flavour fractions
calculated for each variation and the central value was taken
as a systematic error on the determined flavour fractions. 

\subsubsection{Detector Modelling Uncertainties}

\begin{figure}[ht]
\begin{center}
\resizebox{\textwidth}{!}{\includegraphics{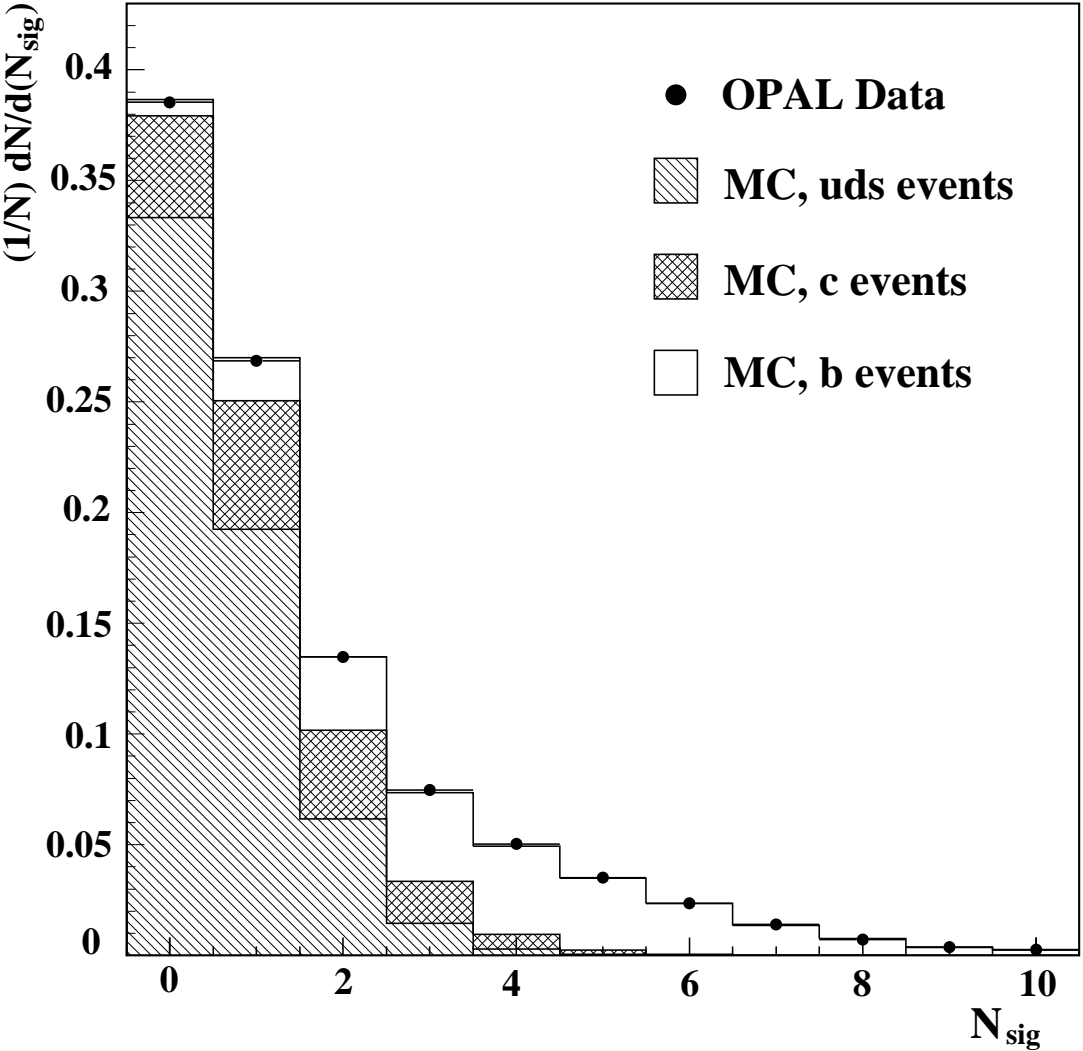}}
\caption{ Distribution of number of tracks per event (N$_{sig}$) with 
$b/\sigma_b > 2.5$. The points are data and the solid histogram
is Monte Carlo. The Monte Carlo is also shown decomposed into 
contributions from uds, c and b quark events. }
\label{fig:nsig}
\end{center} \end{figure}

  Since the determination of the flavour fractions of the uds-tag
and the b-tag event sample depend on Monte Carlo this requires
an accurate simulation of the detector resolution for charged
tracks measured with the silicon micro-vertex detector. 
The simulation has been tuned to reproduce the
tracking resolutions seen in data by studying the impact parameter
distributions of tracks, as functions of track momentum,
polar angle and the different sub-detectors contributing
hits. This tuning procedure was affected by uncertainties in 
the radial alignment within the silicon micro-vertex detector, the
efficiency for associating silicon hits to tracks and the modelling
of known inefficient regions, and  the overall track reconstruction efficiency.
These uncertainties were evaluated according to the procedure given in
\cite{opalrb} and their effect on the flavour fractions determined 
from Monte Carlo was calculated.
In addition, the agreement between data and Monte Carlo in 
Fig.~\ref{fig:nsig} was improved by degrading the resolution of 
impact parameters in the Monte Carlo simulation
by 5\%. This was done by applying a single multiplicative
factor $\beta$ to the difference between the reconstructed 
and true impact parameters. 
Fig.~\ref{fig:nsig} is shown with this smearing applied.
To evaluate the sensitivity of the 
determined flavour fractions to the tracking resolution the
Monte Carlo smearing was removed.                     

\subsubsection{Physics Modelling Uncertainties}

The tagging efficiencies of the uds and b tags for the various quark flavours
are also sensitive to various physics input parameters in the 
Monte Carlo simulation. The rate of gluon splitting to \ccbar\ was varied
in the range $(2.38\pm 0.48)\times 10^{-2}$, based on the OPAL measurement
\cite{opal_glue_split}, and the rate of gluon splitting to \bbbar\ was varied in the
range $(3.1\pm 1.1)\times 10^{-3}$, based on theoretical expectation 
\cite{glue_split_calc}. The error due to these two sources was negligible.
The production fractions of the different weakly decaying
b hadrons were varied according to the experimental uncertainties
\cite{PDG}. The production fractions of the weakly decaying c hadrons,
the fragmentation of b and c quarks, and the charged decay multiplicities 
and lifetimes of b and c hadrons were varied according to the
prescription  given in \cite{opalrb}. The largest effect on the tag flavour 
fractions comes from the b hadron charged decay multiplicity. All these
uncertainties, along with the uncertainties
due to the detector modelling, were added in quadrature and are given as 
the errors on the flavour fraction shown in Table~\ref{tab:flavor}.

\section{Selection of c quark events}
\label{sec:dstar}

  Events having a high probability to
have originated from \ccbar\ were identified by the presence
of a highly energetic D$^{*+}$ meson\footnote{Throughout this paper
charged conjugate modes are always implicitly included.}. 
These events will be referred to in the text below as the ``c-tag''
event sample.
For the D$^{*+}$ reconstruction and the determination
of the flavour composition of the c-tag event sample 
the methods of a previous
OPAL study have been used~\cite{dstarsel}.          
In brief, five D$^{*+}$ decay modes were searched for:

\begin{center}
\begin{tabbing}
  \hspace{5cm} \= \hspace{5cm} \= \kill
  \> ${\rm D^{*+}} \rightarrow {\rm D^0}\pi^+$ \\
  \> $\phantom{D^{*+} \rightarrow }\hspace{4pt}
       \downto {\rm K^-}\pi^+$                   \>``3 prong''\ ,\\
  \> $\phantom{D^{*+} \rightarrow }\hspace{4pt}
       \downto {\rm K^-}\pi^+\pi^0$              \> ``satellite''\ ,\\
  \> $\phantom{D^{*+} \rightarrow }\hspace{4pt}
       \downto {\rm K^-}\pi^+\pi^-\pi^+$         \>``5 prong''\ ,\\
  \> $\phantom{D^{*+} \rightarrow }\hspace{4pt}
       \downto {\rm K^-}{\rm e}^+ \nu_{{\rm e}}$ \>``electron''\ ,\\
  \> $\phantom{D^{*+} \rightarrow }\hspace{4pt}
       \downto {\rm K^-}\mu^+\nu_{\mu}$          \>``muon''\ .\
\end{tabbing}
\label{eq-decaychannels}
\end{center}

\noindent
No attempt was made
to reconstruct the $\pi^0$ in the satellite channel, nor the
neutrino direction or energy in the electron and muon channels.
The last two channels are referred to as ``semileptonic'' channels
in the following text.  A number of tracks
appropriate for the selected channel were
combined to form a D$^0$ candidate and their invariant mass
was calculated.
Candidates were selected if the reconstructed mass lay
within the expected range for that channel. 
After adding a further track
as a possible pion from the D$^{*+}$ decay, the
combined mass was calculated and the candidate was selected
if the mass difference $\Delta M = M_{D^{*+}}-M_{D^0}$
was within certain limits. 
Some of the D$^{*+}$ selection criteria are given in 
Table~\ref{tab:dstarcuts}.

  To reduce the background in the c-tag event sample we required
$x_{{\rm D^{*+}}} > 0.4$, where 
$x_{{\rm D^{*+}}}=E_{{\rm D^{*+}}}^{\mathrm calc}/E_{\rm beam}$ 
is the scaled 
energy\footnote{In this paper any reference to the scaled energy
                $x$ of a D$^{*+}$ candidate is to be understood as 
                being the calculated energy of the D$^{*+}$, 
                $E_{{\rm D^{*+}}}^{\mathrm calc}$, obtained from the 
                reconstructed
                tracks, without correcting for missing or wrongly
                associated tracks, divided by the beam energy 
                $E_{\mathrm beam}$.}
of the D$^{*+}$ meason. For the 5-prong event selection we required
$x_{{\rm D^{*+}}} > 0.5$. This cut is effective in rejecting
D$^{*+}$ mesons which originate from cascade decays 
of B hadrons and from events where a gluon splits into
a pair of charm quarks.

\begin{figure}[ht]
\begin{center}
\resizebox{\textwidth}{!}{\includegraphics{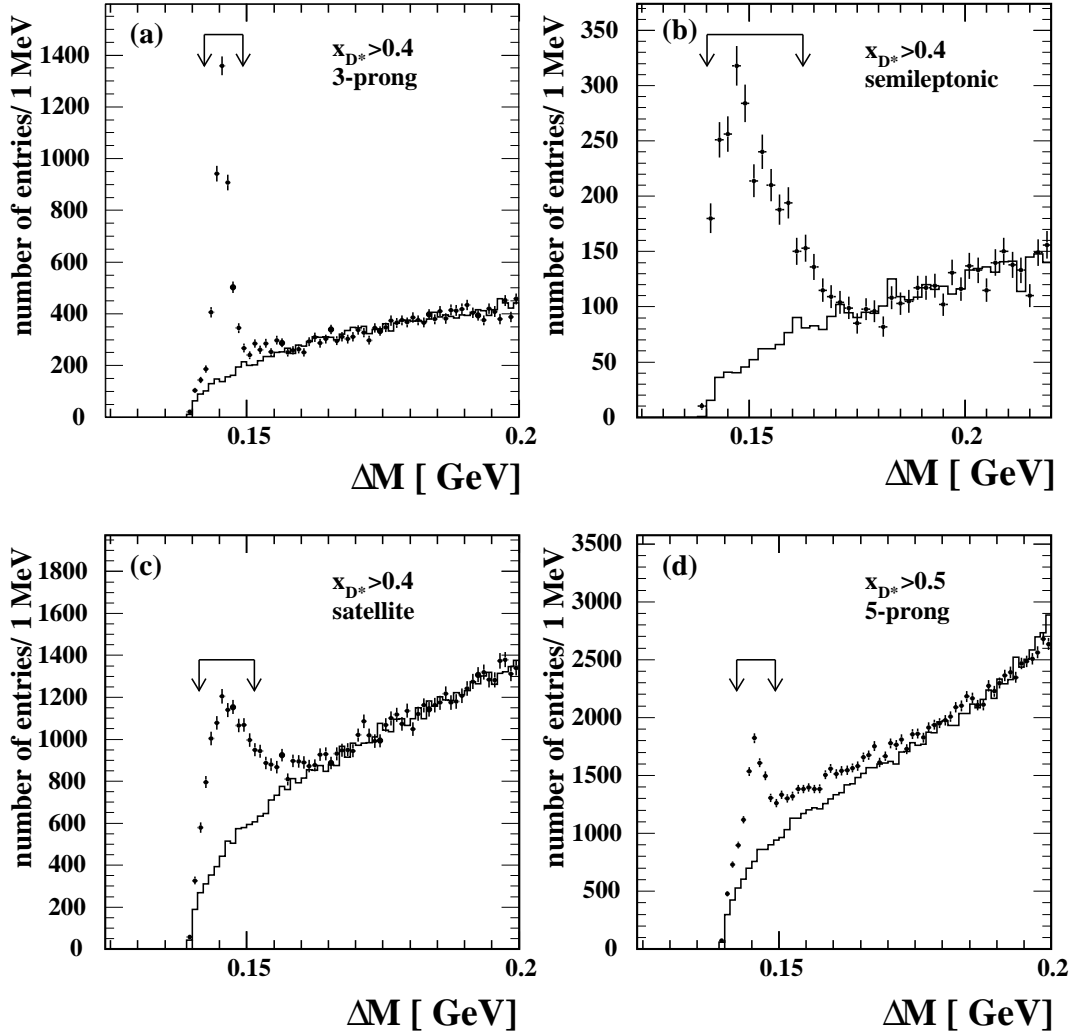}}
\caption{Distribution of the mass difference $\Delta M$ for
four different D$^{*+}$ decay channels: (a) 3-prong decay mode,
(b) the two semileptonic modes combined, (c) the satellite decay mode, 
and (d) the 5-prong decay mode.
The arrows indicate the selected signal regions. 
The points with error bars are the signal candidates and the solid
histograms are the background estimator distributions.}
\label{fig:delm}
\end{center} \end{figure}

  The distribution of $\Delta M$ for all five decay channels
is presented in Fig.~\ref{fig:delm}. The points with error bars
are the signal candidates and the solid histograms are
background estimator distributions constructed from data.
The background estimator was constructed by choosing the
candidate for the pion in the 
$ {\rm D^{*+}} \rightarrow {\rm D^0}\pi^+$
decay from the opposite hemisphere relative to the rest 
of the decay products, reflecting it through the orgin, 
and then using it in the calculation of the invariant mass.

  A significant fraction of the sample of D$^{*+}$ mesons were only partially
reconstructed. These mesons produce an enhancement in the
$\Delta M$ spectrum very similar to the true signal. Only a few
of the events are present in the 3-prong sample. They are more important in
the 5-prong tagged events where a clear tail is visible in the $\Delta M$
distribution for values above 0.145 GeV (see Fig.~\ref{fig:delm}d). 
Since such events originate from D$^{*+}$ decays, they can still be used in 
the analysis.

  In all, 27\,005 D$^{*+}$ candidate events were selected and the
background was estimated to be $11\,366 \pm 107$ events, where
the error on the number of background events is 
statistical only. All of the LEP-1 data recorded by OPAL,
4.4 million events, was used for the D$^{*+}$ event selection.
The candidate events are composed of three 
components: genuine D$^{*+}$ mesons from b events, 
genuine D$^{*+}$ mesons from c events and combinatorial
background which is a mixture of uds, c and b quark events. 
Genuine D$^{*+}$'s from uds quark events can only occur in
events where a gluon splits into two heavy quarks. 
The possibility of D$^{*+}$'s from this source was neglected because
these events are highly suppressed due to the 
high $x_{\rm D^{*+}}$ cut. Using Monte Carlo simulation
the total contribution of this source to the number     
of D$^{*+}$ candidates was found to be $(0.2 \pm 0.1)$\%, where the
error quoted is due to Monte Carlo statistics.

  In Ref.~\cite{dstarsel} the fraction of genuine D$^{*+}$
mesons originating from c events was determined by OPAL to be 
$f_{\rm c}^{{\rm D^{*+}}} = 0.774 \pm 0.023$, where the error is the
combined statistical and systematic error. The fraction of 
genuine D$^{*+}$ mesons from b events is given by
$f_{\rm b}^{{\rm D^{*+}}} = 1 - f_{\rm c}^{{\rm D^{*+}}}$. 

The fractions of uds, c and 
b quark events in the combinatorial background, as determined
from Monte Carlo, were
$0.584 \pm 0.009$, $0.238 \pm 0.009$ and
$0.178 \pm 0.009$, respectively. The errors are a combination
of statistical errors due to finite Monte Carlo statistics
and a systematic error accounting for the
overall quality of the background estimation
procedure. Based on a Monte Carlo study~\cite{dstarsel}
the background estimate was found to be accurate to within 1\%,
and an additional 1\% error was therefore
assigned to the number of background events.
Since that study was flavour blind, a conservative approach was
taken here where the entire error was assigned in turn to each
flavour component in the combinatorial background.

  The overall flavour composition of the c-tag event sample is 
presented in Table~\ref{tab:flavor} along with the combined
statistical and systematic errors. The efficiency for tagging
c events was about 2.0\%.

\section{Event shape observables}
\label{sec:shp}

  For each flavour tagged event sample described
above (uds-tag, c-tag and b-tag)
the distributions of the event shape variables
1-Thrust ($1-T$), Heavy Jet Mass scaled by the
centre-of-mass energy ($M_H/\sqrt{s}$),
Wide Jet Broadening ($B_W$),
the $y_{cut}$ at which an event changes from
being a 2-jet event to being a 3-jet event
($y_{23}$) determined using the Durham jet 
finder and the C-parameter ($C$) were determined. 
The definitions for
these observables are given in~\cite{opalresummed, as_global} 
and the references therein. 
These quantities were measured using all tracks and
electromagnetic clusters which satisfied the 
selection criteria described in Sect.~\ref{sec:evt}.

  In order to extract the values of \rascasu\ and
\rasbasu , the measured event shape distributions were
fitted using a QCD analytic calculation~\cite{heavyQ3}.
Since the QCD calculation is only valid for event shape distributions
determined from final-state partons, hadronization effects
caused by the transformation of final-state partons into hadrons, which are
experimentally accessible,
must be taken into account. This was done
by applying correction factors to the analytic predictions.

  The measured event shape distributions must be corrrected for
experimental effects which distort them.
These effects include finite detector resolution,  
initial-state photon radiation and biases introduced by
the flavour tagging methods.
In the text below describing the correction procedure
the term {\it detector level} is used to refer
to distributions determined using the measured tracks and
electromagnetic clusters and {\it hadron level} to refer to
these distributions corrected for detector resolution,
initial state radiation and biases introduced by the flavour
tagging methods. The comparisons between the measured 
distributions and the QCD predictions were performed
at the hadron level.

The event shape variable Total Jet Broadening 
($B_T$)~\cite{opalresummed}
was initially considered to be included 
in this analysis. It was found, however, that for the distribution of $B_T$
measured with b quark events, the size of the hadronization
corrections were greater that 20\%  
over the entire distribution. Therefore
this variable was dropped from further consideration.

\subsection{Correction procedure} 
\label{sec:cor}

  In the first step the selected events     
were corrected for distortions caused by finite 
detector resolution using an unfolding matrix. 
The unfolding matrix was constructed using
a Monte Carlo data sample including 
initial-state radiation, full detector
simulation and subjected 
to the same event selection criteria that were applied
to the data. The Monte Carlo events which pass the
selection criteria were used to calculate a 
correction matrix $M^{q-tag}$. 
The element $M^{q-tag}(y_i,y_j)$ gives the 
probability that an event shape observable
$y$ measured at the detector level and located in
bin $i$ of its corresponding distribution,
has migrated from bin $j$ on the hadron level.
A matrix was computed
for each event shape observable and flavour-tagged
event sample,
``q-tag'', where q-tag was either uds-tag, c-tag or b-tag.

  In the next step bin-by-bin corrections are used
to correct the data for biases introduced 
by the flavour tagging methods, event acceptance and 
the effects of initial-state photon radiation. 
Defining $G^q(y_i)$ to be the number of events of
flavour $q$ in
the untagged Monte Carlo sample and $H^{q-tag}(y_i)$
the number of events with the tag applied,
the correction factor $K^{q-tag}(y_i)$ for the $i$-th bin is 
given by

\begin{eqnarray}
 K^{q-tag}(y_i) = 
                   \frac{f^{q-tag}_{\rm uds} G^{\rm{uds}}(y_i) +
                         f^{q-tag}_{\rm   c} G^{\rm{  c}}(y_i) +
                         f^{q-tag}_{\rm   b} G^{\rm{  b}}(y_i)  }
                        {H^{q-tag}(y_i)}
\end{eqnarray}

\noindent
The factors $f_q^{q-tag}$, taken from 
Table~\ref{tab:flavor},
are the fractions of events of flavour $q$ in each flavour-tagged      
event sample. The terms $G^q(y_i)$ are normalized to the total
number of events of flavour $q$.

\begin{figure}[ht]
\begin{center}
\resizebox{\textwidth}{!}{\includegraphics{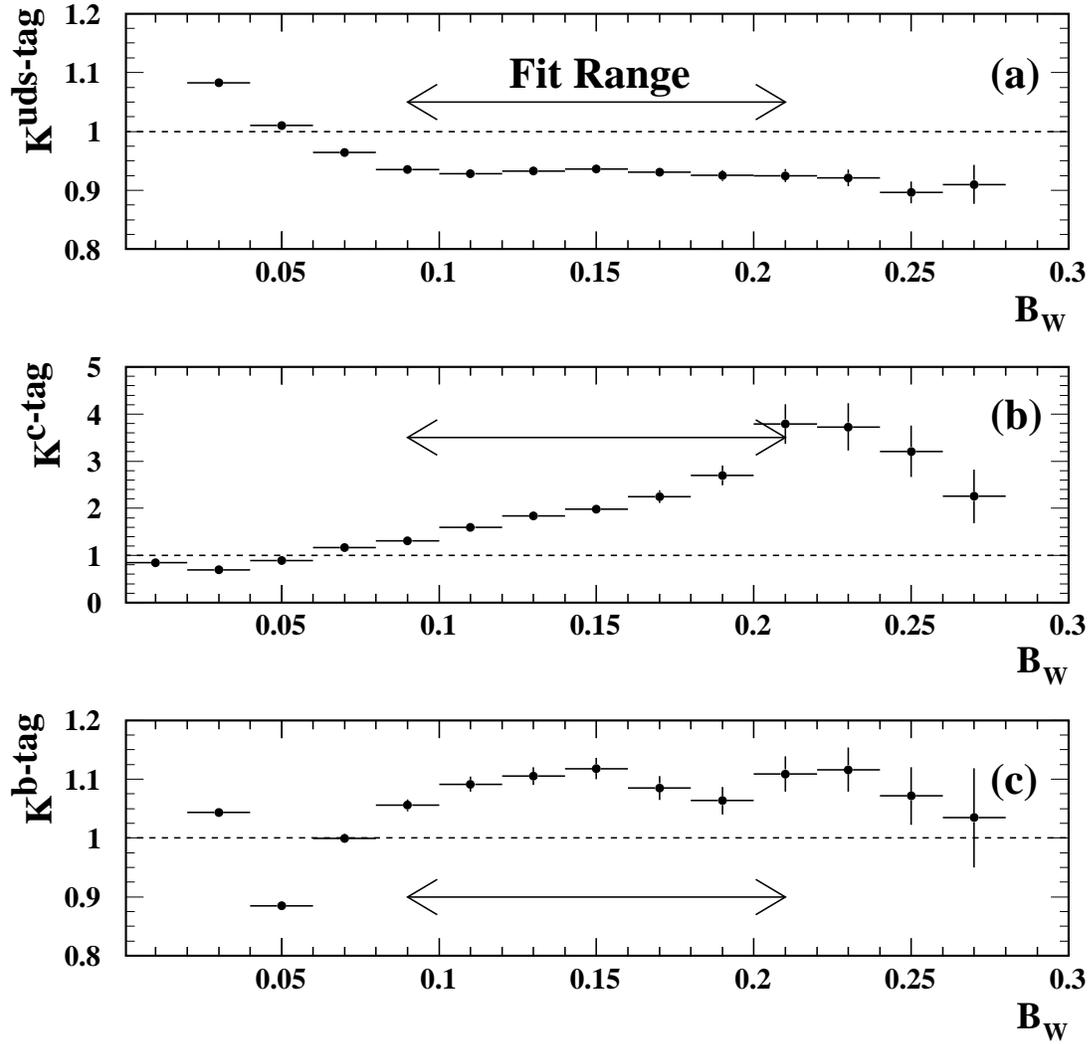}}
\caption{Bin-by-bin correction factors applied to the distribution of $B_W$ for
         (a) the uds-tag event sample, (b) the c-tag event sample and
         (c) the b-tag event sample. The fit range for this observable 
         (see Sect.~\ref{sec:oas2}) is indicated by the arrow on each figure.} 
   \label{fig:bwcor}
\end{center} \end{figure}

  Fig.~\ref{fig:bwcor} shows, as an example, the size of the bin-by-bin corrections
$K^{q-tag}$
for the observable $B_W$ measured using each flavour-tagged
event sample. The size of the corrections for 
$B_W$ were typical of the other event shape observables studied.
One can see for the distributions
measured with the uds-tag and b-tag event samples that, within the 
chosen fit range, the size of the corrections are about
10\%. (An explanation of how the fit range was chosen is presented in
the next section.) In contrast, the size of the corrections for the 
c-tag event sample were an order of magnitude larger. This is due 
to a kinematic
bias introduced by requiring the events in the 
c-tag sample to contain a high-$x$ $D^{*+}$ meson. 

  The number of events in bin $i$ corrected to the hadron level is
given by

\begin{eqnarray}
N^{q-tag,cor}(y_i) = K^{q-tag}(y_i) \sum_{j} M^{q-tag}(y_i,y_j)N^{q-tag}(y_j)~~,
\end{eqnarray}

\noindent
where $N^{q-tag}(y_j)$ is the uncorrected number of events
in the $j$-th bin of the event shape distributions in question.
This correction procedure does not depend on
the values of \asu , \rascasu\ or \rasbasu\ in the 
Monte Carlo samples.

The resulting distributions normalized to the total hadronic
cross section are given by

\begin{eqnarray}
\left(\frac{1}{\sigma_{tot}}\frac{\rm{d}\sigma}{\rm{d}y}\right)^{q-tag,cor}
= \frac{1}{\Delta y_i \cdot \sum_{j} N^{q-tag,cor}(y_j)} N^{q-tag,cor}(y_i)~~,
\end{eqnarray}

\noindent
where $\Delta y_i$ is the width of the $i$-th bin and $y$ corresponds
to one of the five event shapes studied. 

\section{Fit procedure}                                   
\label{sec:oas2}

  The value of \asu\ and the ratios \rascasu\ and
\rasbasu\ were determined by simultaneously fitting theoretical
predictions for a particular event shape observable to three
hadron-level
event shape distributions: one determined with the uds-tag event sample,
one with the c-tag event sample and one with the b-tag event sample.
The theoretical predictions were given by a linear combination of the 
theoretical predictions for each of the three tagged flavours: 

\begin{eqnarray}
\label{eq:theo_shp}
 \left(\frac{1}{\sigma_{tot}}\frac{\rm{d}\sigma}{\rm{d}y}\right)^{q-tag,th} &=& 
\label{eqn:lincomb}
      f_{\rm{uds}}^{q-tag}R(y)^{\rm{uds}}\left(\frac{1}{\sigma_{tot}}\frac{\rm{d}\sigma}{\rm{d}y}\right)^{{\rm uds},th}   \nonumber
  +   f_{\rm{  c}}^{q-tag}R(y)^{\rm{c  }}\left(\frac{1}{\sigma_{tot}}\frac{\rm{d}\sigma}{\rm{d}y}\right)^{{\rm c  },th} \\
&  & \mbox{} + f_{\rm{  b}}^{q-tag}R(y)^{\rm{b  }}\left(\frac{1}{\sigma_{tot}}\frac{\rm{d}\sigma}{\rm{d}y}\right)^{{\rm b  },th}~~. 
\end{eqnarray}

\noindent
The coefficients $f_q^{q-tag}$ are the flavour fractions
given in Table~\ref{tab:flavor} and 
($1/\sigma_{tot}\cdot {\rm d}\sigma/{\rm d}y)^{\rm{uds},th}$,
($1/\sigma_{tot}\cdot {\rm d}\sigma/{\rm d}y)^{\rm{c},th}$ and
($1/\sigma_{tot}\cdot {\rm d}\sigma/{\rm d}y)^{\rm{b},th}$ 
are the theoretical predictions for an event shape observable
$y$ measured with a sample of uds, c and b quark events, 
respectively. 
The factors $R(y)^q$ correct
the theoretical prediction for 
a particular tagged flavour $q$ for hadronization 
effects so that the theoretical predictions, valid for
final state partons, can be compared
directly with the measured distributions 
corrected to the hadron level. 

  The hadronization correction factors $R(y)^q$ were computed
from JETSET 7.4 using the 
parton shower option
by taking the ratio of an event shape distribution at the
hadron level to the same distribution at the parton
level. Here the parton level is defined by the cut-off $Q_0$ of
the QCD shower in JETSET which is set to 1.9\,GeV~\cite{bib-OPALtune}. 
After the termination of the parton shower the partons
are transformed into hadrons using string 
hadronization~\cite{string-had}. For this process
we choose a hybrid scheme for the longitudinal fragmentation 
function where light quarks are treated with the symmetric
Lund fragmentation function and charm and bottom quarks according to the 
model of Peterson et al.~\cite{bib-PETERSON}.
The c and b quark masses in JETSET were set to their
default values of 1.35\,GeV and 5.0\,GeV, respectively.

In Equation~(\ref{eqn:lincomb}),
the differential cross section of a generic observable $y$,
for massless quarks,
normalized to the total hadronic cross section is
given by~\cite{oas2}


\begin{eqnarray}
\left(\frac{1}{\sigma_{tot}}\frac{{\rm d}\sigma}{{\rm d}y}\right)^{{\rm uds},th} & = &\nonumber
\label{eqn:udsth}
  \frac{{\rm d}A^{\rm uds}}{{\rm d}y}\left(\frac{\alpha_s^{uds}(\mu)}{2\pi}\right) \\ 
  &  & \mbox{} + \left(\left(2\pi \beta_0 \mathrm{log}(x_\mu^2) - 2 \right)
  \frac{{\rm d}A^{\rm uds}}{{\rm d}y} + \frac{{\rm d}B^{\rm uds}}{{\rm d}y}\right)
  \left(\frac{\alpha_s^{uds}(\mu)}{2\pi}\right)^2~~. 
\end{eqnarray}

\noindent
The coefficients
${\rm d}A^{\rm uds}/{\rm d}y$ and ${\rm d}B^{\rm uds}/{\rm d}y$ are the 
\oa\ and \oaa\ QCD coefficients,
respectively, $\sigma_{tot}$ is the one-loop cross section
for the process $\epem \rightarrow \mathrm{hadrons}$ and
$\beta_0$ is the coefficient of the QCD beta function
for one-loop~\cite{PDG}. The renormalization scale
$\mu$ can be related to the \epem\ centre-of-mass
energy by

\begin{eqnarray}
\mu = x_\mu \cdot E_{cm}~~,
\end{eqnarray}

\noindent
where $x_\mu$ is the renormalization scale factor.
The coefficients ${\rm d}A^{\rm uds}/{\rm d}y$ and ${\rm d}B^{\rm uds}/{\rm d}y$
were obtained for each event shape observable 
by integrating the \oaa\ matrix elements in~\cite{oas2}.
In Equation~(\ref{eqn:udsth}), in addition to the approximation
of massless quarks,
the simplifying assumption was made that \as\ is the same for
up, down and strange quarks. 

The terms
($1/\sigma_{tot}\cdot {\rm d}\sigma/{\rm d}y)^{\rm{c},th}$ and
($1/\sigma_{tot}\cdot {\rm d}\sigma/{\rm d}y)^{\rm{b},th}$
in Equation~(\ref{eqn:lincomb}) were both given by an 
\oaa\ expression for massive quarks.
This expression has only recently been made available~\cite{heavyQ3}.
The result of this calculation was implemented in a FORTRAN program
named ZBB4~\cite{zbb4} analogous to the program
EVENT~\cite{event} which was used to integrate the \oaa\
matrix elements of the massless calculation. 
ZBB4 was 
run separately for c quark and b quark events in order
to calculate the \oa\ and \oaa\ coefficients for
the massive calculation. This calculation was performed
in the pole mass scheme and
the c and b quark pole masses
were set to 1.35\,GeV and 5.0\,GeV, respectively.

The differential cross section of a generic observable $y$,
for massive quarks, normalized to the total hadronic cross section is
given by~\cite{zbb4} 


\begin{eqnarray}
\left(\frac{1}{\sigma_{tot}}\frac{{\rm d}\sigma}{{\rm d}y}\right)^{Q,th} & = &  \nonumber
\label{eqn:Qth}
  \frac{{\rm d}A^{Q}}{{\rm d}y}\left(\frac{\alpha_s^{Q}(\mu)}{2\pi}\right) \\ 
  &   & \mbox{} + \left(\left(2\pi \beta_0 \mathrm{log}(x_\mu^2) - 
  \frac{2}{3}{\mathrm{log}}\left(\frac{\mu}{m_{Q}}\right) - 2 \right)
  \frac{{\rm d}A^{Q}}{{\rm d}y} + \frac{{\rm d}B^{Q}}{{\rm d}y}\right)
  \left(\frac{\alpha_s^{Q}(\mu)}{2\pi}\right)^2
\end{eqnarray}

\noindent
where $Q$ is either c or b and $m_Q$ is the energy scale corresponding 
to the heavy quark pole mass.

  When performing the fit using Equation~(\ref{eqn:lincomb}) we make
the substitution $\asc~=~\asu \cdot \rascasu$ and 
$\asb =~\asu \cdot \rasbasu$. This substitution enables
\asu\ , \rascasu\ and \rasbasu\ to be determined
directly as free parameters in the fit and allows
correlations between these variables to be properly taken
into account.
In tests with Monte Carlo events it was verified
that this fitting procedure was sensitive to changes in \as\ for 
c and b quarks with respect to uds quarks.

  The fit ranges for each observable were determined by the range 
of $y$ where the parton to hadron level corrections were below
10\% and the resulting $\chi^2/\mathrm{d.o.f.}$ remained
small ($\sim 5$ for $B_W$ and $\sim 2-3$ for the rest).
The remaining variations of the fit
results due to the choice of the fit ranges were taken as
systematic uncertainties. It was also checked that 
within the fit range the \oaa\ calculation and the JETSET parton shower 
model were in agreement with each other. The fit ranges used for this analysis
are slightly smaller than the fit ranges used in previous OPAL
studies~\cite{opalresummed, as_global} which used \oaa\ calculations in fits
to global event shapes. This is because the hadronization corrections for
distributions of event shape variables calculated with
b quark events are generally larger than the
corrections for distributions calculated with an inclusive event 
sample. Consequently, the range of the distribution
in which the corrections are below 10\% is smaller.

  The fit ranges, and
the results of the $\chi^2$ fits with $x_\mu = 1$ are shown in 
Table~\ref{tab:oas2}.
The values for \asu\ obtained appear 
large when compared with \as\ measurements performed using
resummed NLO calculations~\cite{opalresummed}.
This is due to the fact that fixed order 
QCD calculations, valid only to \oaa , were used here. 

  To estimate the error in our results due to the choice of the
renormalization scale we adopted a method used in another
OPAL analysis~\cite{as_global} which used \oaa\ fits to global event
shapes. We performed fits with $x_\mu$ fixed equal one, 
and fits in which $x_\mu$ was an additional free parameter. The average of
the two results is taken as our main result, and half
their difference as the ``scale uncertainty''.
The result of fits with $x_\mu$ as a free parameter are presented
in Table~\ref{tab:oas2free}. One sees that the values of
\asu\ decrease for all event shapes
and that the values of $\chi^2/$d.o.f. show a significant improvement. 
This strong scale dependence is typical of determinations of \as\ where fixed
order calculations are used. It was checked that the values of \asu\
determined here were consistent with previous OPAL measurements of 
\as~\cite{opalresummed,as_global}. The possibility of a different
scale dependence for heavy quark events, due to
mass effects present in higher order terms, was 
investigated by introducing a separate $x_\mu$ parameter for each 
quark flavour. The effect was found to be negligible within the 
statistical precision of the measurement.

\begin{figure}[ht]
\resizebox{\textwidth}{!}{\includegraphics{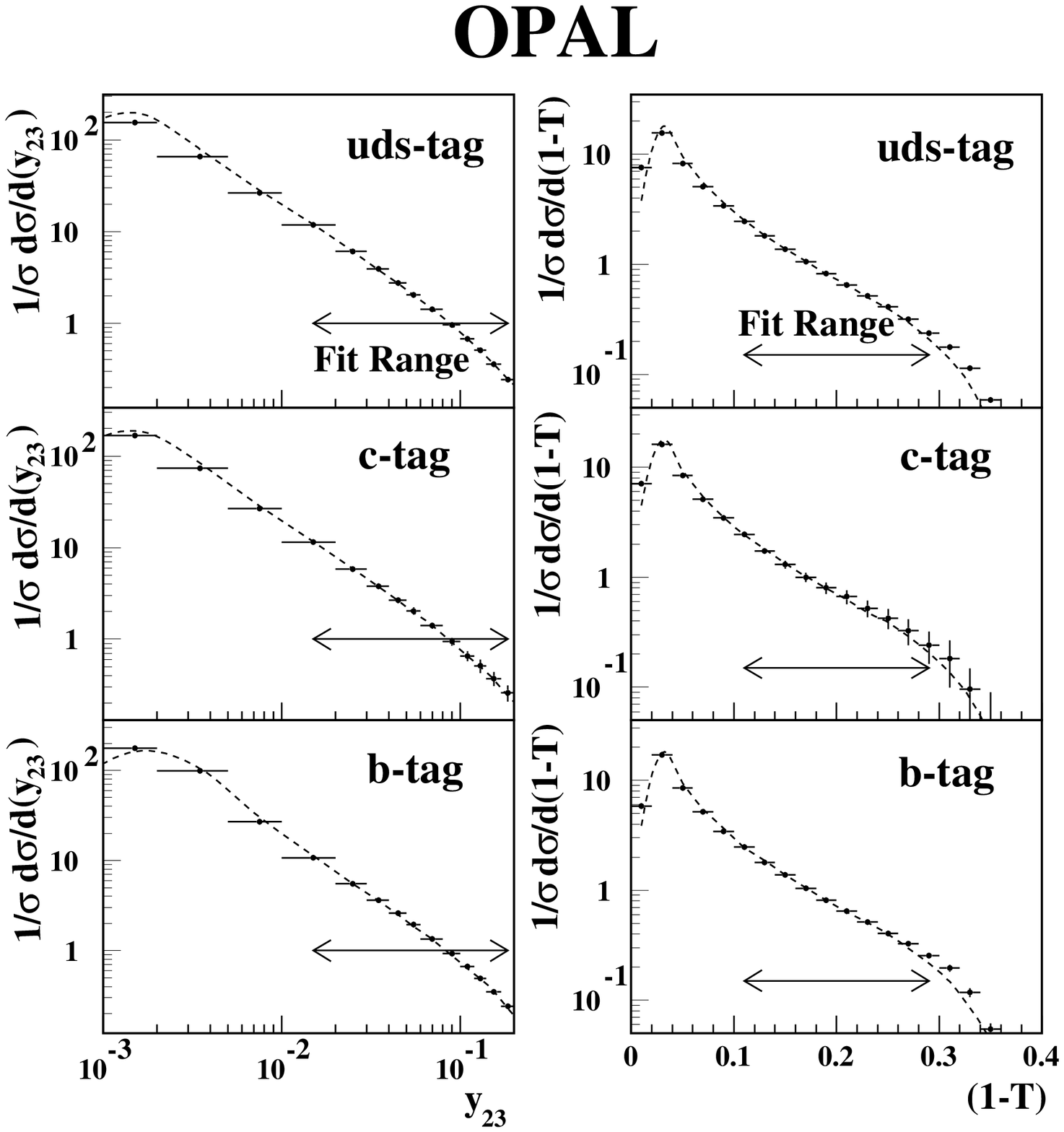}}
\caption{The points with error bars are the 
distributions of $y_{23}$ and $1-T$ measured in
three flavour-tagged event samples (uds-tag, c-tag and b-tag) 
corrected to the hadron level. The curves show the result of         
simultaneous fits of \oaa\ calculations to the distributions
measured in each flavour-tagged event sample. 
The renormalization scale $x_\mu$ was fixed equal to one.
The fit range for each
distribution is indicated by the arrows.}
   \label{fig:d2th}
\end{figure}

\begin{figure}[ht]
\resizebox{\textwidth}{!}{\includegraphics{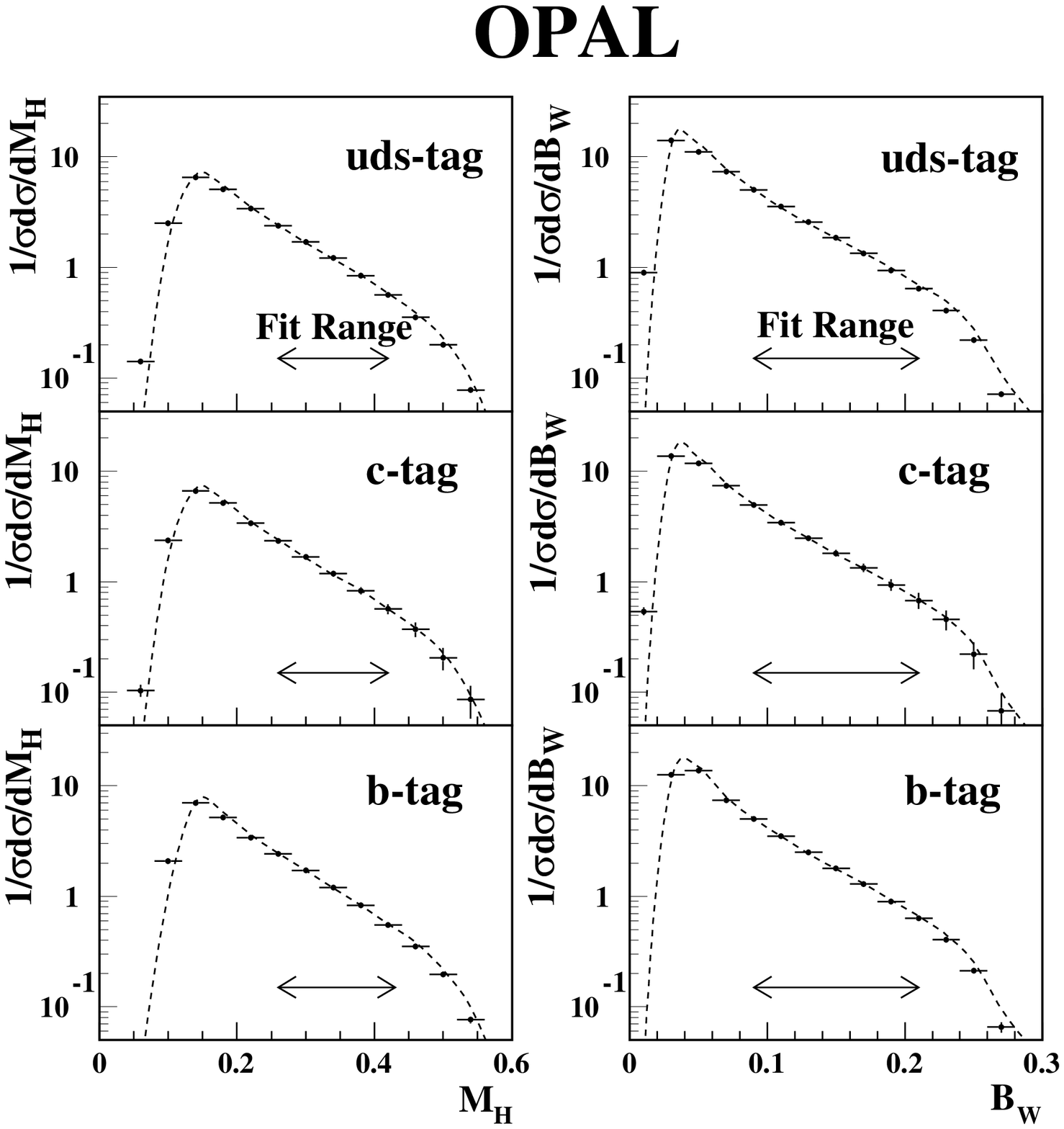}}
\caption{The points with error bars are the 
distributions of $M_H$ and $B_W$ measured in
three flavour-tagged event samples (uds-tag, c-tag and b-tag) 
corrected to the hadron level. The curves show the result of         
simultaneous fits of \oaa\ calculations to the distributions
measured in each flavour-tagged event sample. 
The renormalization scale $x_\mu$ was fixed equal to one.
The fit range for each
distribution is indicated by the arrows.}
   \label{fig:mhbw}
\end{figure}

\begin{figure}[ht]
\begin{center}
\resizebox{.55\textwidth}{!}{\includegraphics{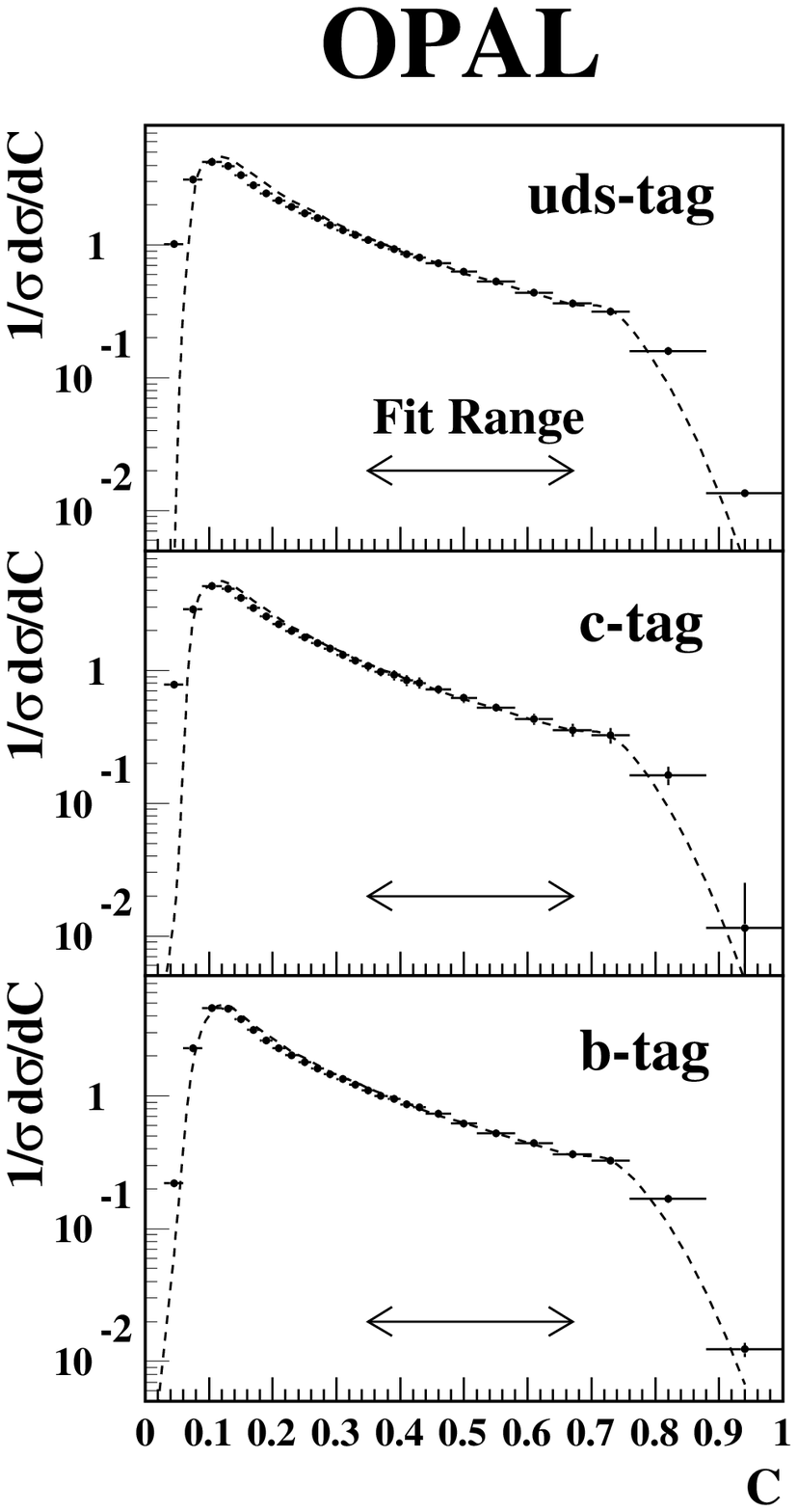}}
\caption{The points with error bars are the 
distribution of the $C$ parameter measured in
three flavour-tagged event samples (uds-tag, c-tag and b-tag) 
corrected to the hadron level. The curves show the result of         
simultaneous fits of \oaa\ calculations to the distributions
measured in each flavour-tagged event sample. 
The renormalization scale $x_\mu$ was fixed equal to one.
The fit range for each
distribution is indicated by the arrows.}
   \label{fig:cp}
\end{center}
\end{figure}

  In Figs.~\ref{fig:d2th} through~\ref{fig:cp} the results of
fits with $x_\mu = 1$ are
plotted along with the measured distributions corrected to
the hadron level. The fit ranges used are indicated by the
arrow on each figure. It can be seen that within the fit ranges
the agreement between the fit results and the data are good. 
In Fig.~\ref{fig:d2th} one sees a small disagreement
between the fit results and the data for values of $1-T > 0.25$.
When the fits to $1-T$ were repeated with $x_\mu$ as a free parameter,
the agreement between the fit results and data were
good over the entire fit range.

\section{Systematic uncertainties}
\label{sec:sys}

The main result was obtained
using the default selection and correction procedure described above.
The systematic uncertainties were divided into two groups:
experimental and theoretical uncertainties. 
Each uncertainty was estimated by modifying details
of the event selection and correction procedure and
repeating the analysis. The difference between
the results obtained with the standard analysis and the
results obtained with the
analysis corresponding to each variation were taken (unless otherwise
noted) as symmetric systematic errors. In the case where a 
parameter was varied above or below its nominal value, the
largest deviation from the main result was taken as a
symmetric systematic error associated with that parameter.
Finally, the systematic errors for each type
of variation were added in quadrature. The systematic 
uncertainties investigated are described below and the effects they 
had on the ratios \rascasu\ and \rasbasu\ are presented
in Tables~\ref{tab:sys_ascasu} and \ref{tab:sys_asbasu}.

\subsection{Experimental systematic uncertainties}

\begin{itemize}

\item The error due to the uncertainty on the flavour composition
of the flavour-tagged event samples was evaluated by varying the 
flavour fractions in Table~\ref{tab:flavor} within their errors.

\item For the central result event shape observables 
were measured using all 
charged tracks and electromagnetic clusters.
To evaluate the relative response of the central tracking and  
the electromagnetic calorimeter the measurements were performed
again using electromagnetic clusters only and charged tracks only.                            

\item The homogeneity of the response of the detector in the endcap region
was checked by restricting the analysis
to the barrel region of the
detector, requiring the thrust axis of accepted events to lie within
the range $|\cos\theta_{\rm Th}| < 0.7$. 
This only modifies event shape distributions measured with
the c-tag event sample
because the uds-tag and b-tag event samples are already restricted
to the barrel region.

\item The minimum number of accepted charged tracks was
increased from 5 to 7 in order to further suppress background from 
$\tau^+ \tau^-$ events and two-photon interactions.

\item The dependency of the result on the chosen fit range was
evaluated by adding or subtracting one bin from the lower and 
upper end of the fit range. Each new fit range was treated as a
separate systematic variation. 

\item In the charm event selection the requirement that 
$x_{D^{*+}} > 0.4$ leads to a large kinematic bias in
the c-tag event sample. The uncertainty
due to this aspect of the $D^{*+}$ selection was evaluated by
varying the $x_{D^{*+}}$ cut between 0.3 and 0.5.

\end{itemize}

\subsection{Theoretical systematic uncertainties}

\begin{itemize}

\item The fragmentation of the final-state partons
into hadrons was modelled by JETSET
where the longitudinal fragmentation function for light
quarks was treated by the symmetric Lund model and
for heavy quarks by the model of Peterson et al~\cite{bib-PETERSON}.
The parameters for each model were determined from 
a fit to OPAL data on global event shapes~\cite{bib-OPALtune}.
This fit yielded a value of $b = 0.52 \pm 0.04$ ({\tt PARJ(42)})
for the Lund model and $\epsilon_c = 0.031 \pm 0.011$ 
({\tt PARJ(54)}) and $\epsilon_b = 0.0038 \pm 0.0010$
({\tt PARJ(55)}) for the Peterson model.
To evaluate the systematic error
due to uncertainties in the JETSET hadronization model these
parameters were varied independently within one standard deviation of 
their optimised values. 

\item In JETSET, the parameter $Q_0$ ({\tt PARJ(82)}) is the virtuality cut-off
of partons. It determines the boundary between the
pertubative QCD and hadronization phases and is essentially  
arbitrary. The optimum value to describe the OPAL data
was determined to be $Q_0 = (1.90 \pm 0.50)$\,GeV~\cite{bib-OPALtune} .
This parameter was varied within one standard deviation of its
optimised value.

\item The width of the transverse momentum distributions of quarks and antiquarks
produced in the fragmentation process is determined by the
parameter $\sigma_q = (0.40 \pm 0.03)$\,GeV ({\tt PARJ(21)}). 
This parameter was varied within one standard deviation of its
optimum value.                                             

\item The renormalization scale uncertainty was estimated by performing
fits with $x_\mu$ fixed equal one, 
and fits in which $x_\mu$ is an additional free parameter. The average of
the two results is taken as our main result, and half
their difference as the scale uncertainty. 

\item To evaluate the dependence of this analysis on the 
choice of using JETSET to calculate the hadronization corrections,
the analysis was repeated using ARIADNE 4.08~\cite{ariadne} instead 
of JETSET. The ARIADNE parton shower
is based upon a colour dipole model and provides an alternative to
the Lund parton shower model in JETSET. ARIADNE employs the same
fragmentation model as JETSET for the subsequent hadronization.
The difference between the results
using JETSET hadronization corrections and ARIADNE corrections were
used as an estimate of this systematic error.

\item The error due to uncertainties in the
c and b quark masses was evaluated by varying the 
quark masses in the ranges given in~\cite{PDG} and repeating      
the integration of the matrix elements corresponding to massive quarks
for each variation. The value for the b quark pole mass was
varied from 4.5 to 5.5\,GeV/$c^2$; 
the c quark pole mass was varied from 1.2 to 1.9\,GeV/$c^2$. 

\end{itemize}

\section{Results}
\label{sec:res}

\begin{figure}[ht]
\resizebox{\textwidth}{!}{\includegraphics{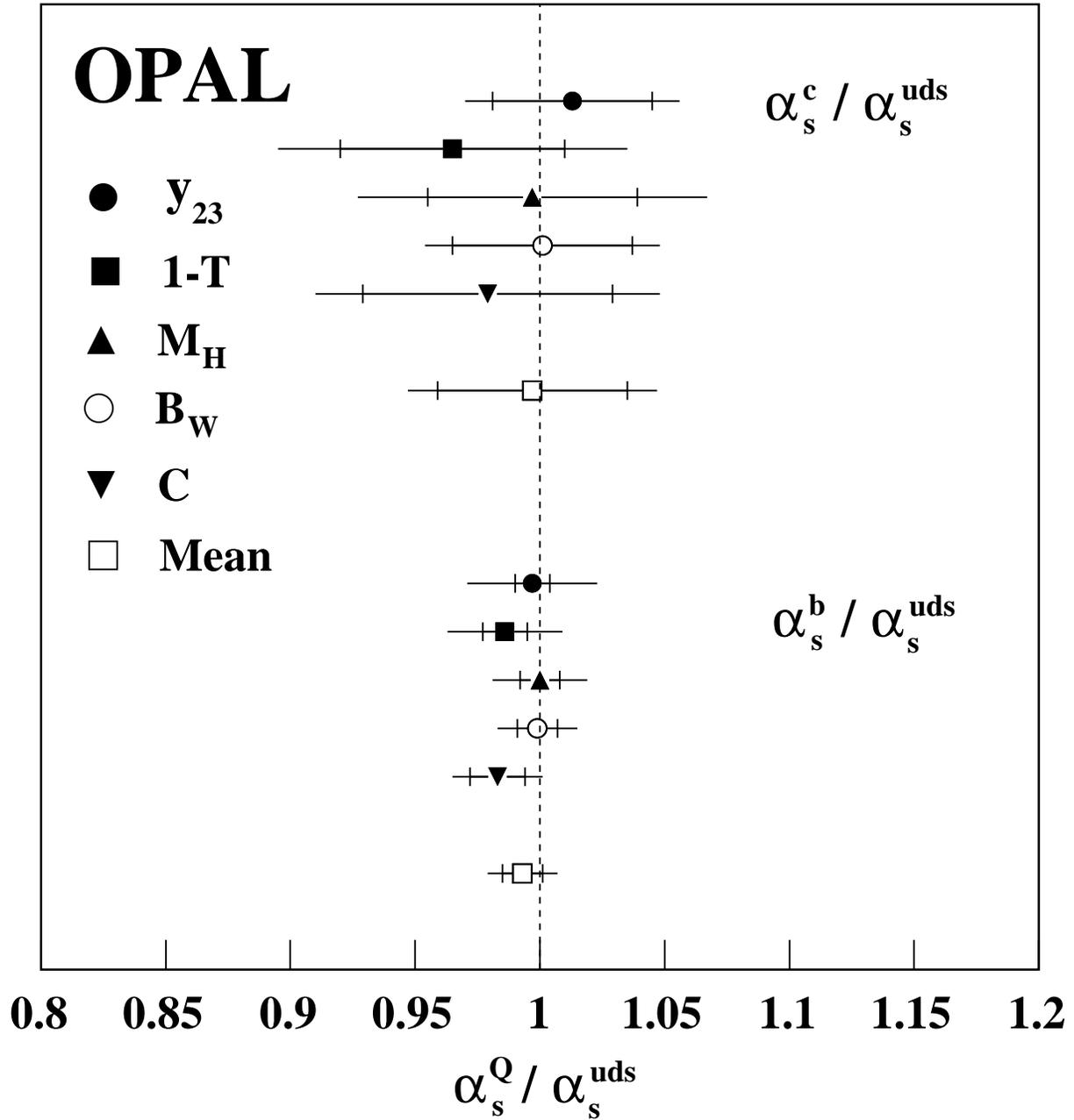}}
\caption{The ratios \rascasu\ and \rasbasu\
determined for each of the event shapes studied.
The weighted mean of the results derived for the five
event shape observables is also shown.
The inner error bar represents the statistical
error while the full error bar represents the statistical error
and the systematic error added in quadrature.}
\label{fig:oas2}
\end{figure}

The results of the $\chi^2$ fits to determine \rascasu\ and
\rasbasu\ for each of the event shape observables studied
are presented in Tables~\ref{tab:sys_ascasu} and \ref{tab:sys_asbasu}
and summarised in Fig.~\ref{fig:oas2}. The values quoted are 
the average of fits with $x_\mu = 1$ and with $x_\mu$ as a free parameter.
The top half of the
Fig.~\ref{fig:oas2} shows the result of determining \rascasu\ and
the lower half \rasbasu . The vertical bars on the error bars
show the size of
the statistical error and the full error bar is the total
error, which is the sum of the
statistical, experimental systematic and theoretical errors
added in quadrature.
Also shown is the weighted mean of the results obtained 
with the six observables. The weights were given by the 
reciprocal of the square of the total error on 
the ratios \rascasu\
and \rasbasu\ given in Tables~\ref{tab:sys_ascasu} and
\ref{tab:sys_asbasu}. The statistical error on the mean
was calculated taking into account correlations between
the five observables. The correlation 
matrix was calculated from 100 Monte Carlo event samples. 
The systematic uncertainty on the weighted mean was
determined from the change in the mean that occurred when each
systematic check was applied to all the event shape distributions
simultaneously.  
The mean values of \rascasu\ and \rasbasu\ 
were determined to be

\begin{eqnarray}
\rascasu &=& 0.997 
  \pm 0.038 ~(stat.) \pm 0.030 ~(syst.) \pm 0.012 ~(theory)  \nonumber \\
\rasbasu &=& 0.993 
  \pm 0.008 ~(stat.) \pm 0.006 ~(syst.) \pm 0.011 ~(theory)~. \nonumber 
\end{eqnarray}

\noindent
In both cases the results are consistent with unity, indicating
flavour independence of \as .

As shown in Table~\ref{tab:sys_asbasu}, the largest experimental systematic error
on the results obtained from the variable $y_{23}$
originated from moving the lower bound of the fit range
from $y_{23} = 0.015$ to 0.025. It was found that this relatively large
variation in the result was due to a small discrepancy between the
data and the Monte Carlo, of the order of a few percent, for values of
$y_{23} < 0.03$. If one excludes the results obtained from $y_{23}$
from the calculation of the weighted means, the mean value obtained for
$\alpha_s^b/\alpha_s^{uds}$ remains unchanged within the experimental
precision, while the mean value of $\alpha_s^c/\alpha_s^{uds}$
changes from 0.997 to 0.990.

The size of the mass effect for c and b quarks 
and its relevance to the measurement of \as\ are also of interest.
From phase space considerations one can estimate the size of the mass 
effects for 3-jet observables. The ratio of the phase space of two massive 
quarks and a gluon to the phase space for three massless particles is 
$1 + 8(\rm{M}_q/\rm{M}_Z)^2 \rm{log}(\rm{M}_q/\rm{M}_Z)$~\cite{rodrigo_thesis}. 
This represents a 7\% effect
for M$_q = 5$\,GeV/$c^2$ and 0.7\% for M$_q = 1.35$\,GeV/$c^2$.

\begin{figure}[ht]
\resizebox{\textwidth}{!}{\includegraphics{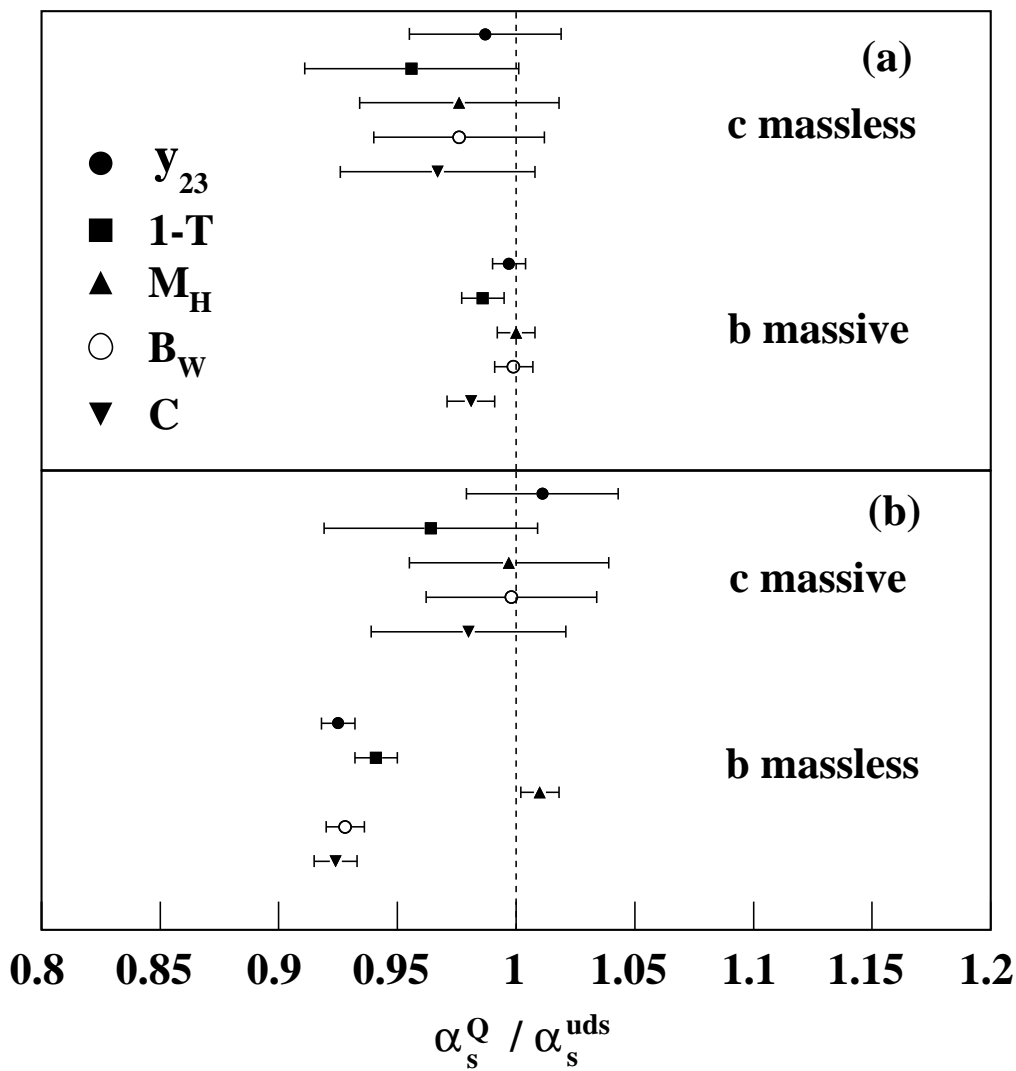}}
\caption{The ratios \rascasu\ and \rasbasu\ determined in two different         
ways: (a) where the uds and c quarks were treated as if massless and the 
b quark was considered to be massive and (b) where the uds and
b quark were treated as massless and the c quark was considered
massive. The errors shown are statistical only.}
\label{fig:masseff}
\end{figure}

Fig.~\ref{fig:masseff}a shows the 
ratios \rascasu\ and \rasbasu\ determined using \oaa\ massless calculations
for both uds and c quarks and \oaa\ massive calculations for b quarks. 
Again we show the average of fits with \xmu\ = 1.0 and \xmu\  
as a free parameter.
One sees a small systematic shift of \rascasu\ with respect to the fits with
the massive calculation, although the
large statistical error on the measurements for c quarks makes a
definitive statement on the exact size of the mass effect difficult. 
In Fig.~\ref{fig:masseff}b,
\rascasu\ and \rasbasu\ were determined using massless calculations for uds and
b quarks and massive calculations for c quarks.
In this case \rasbasu\ shows a large
systematic shift of order 5 to 7\%. 
It should be noted that the values of
\asu\ and \rascasu\ remained unchanged within their statistical errors
with respect to the results presented in Fig.~\ref{fig:oas2}, therefore 
the change in \rasbasu\ can be attributed entirely to changes in \asb .
This sensitivity to the b quark pole mass ( \mb\ ) suggests that this effect 
could be exploited
to measure \mb\ itself, by assuming flavour independence of \as\ and 
fitting for \mb . Monte Carlo studies show that the value of \mb\ determined
in this way depends strongly on the input value of \mb\ used in
JETSET to calculate hadronization corrections, thus making a determination
of \mb\ in this way problematic.

  As expected, fits to $M_H$ do not exhibit  
the mass effect like the other variables do. Here the quark mass
fixes a lower bound of the $M_H$ distribution since the invariant mass of the 
jet cannot be less than the mass of the quark that originates the jet. For 
\mb\ = 5.0\,GeV the lower bound lies at about $M_H/\sqrt{s} = 0.05$,
which is well outside of the fit range.

\section{Conclusion}
\label{sec:con}

  We have presented a test of the flavour independence of the strong
coupling constant for charm and bottom quarks with respect to light 
(uds) quarks. This analysis was based on a sample of hadronic decays 
of the \znull\ resonance recorded by the OPAL detector at LEP.
The global event shapes $y_{23}$, $1-T$, $M_H$, $B_W$ and $C$
were used to measure \as\ in three flavour tagged event samples (uds, c and b). 
The event shape distributions were fitted by \oaa\ calculations
of jet production
taking into account mass effects for the c and b quarks.
The ratios \rascasu\ and \rasbasu\ were both found to be 
consistent with unity, indicating the flavour independence of \as .

The measurement of \rascasu\ achieved a precision of 5\%. The 
relatively large statistical error was due to the low efficiency for
tagging c quark events ($\approx 2\%$) which relied on finding 
D$^{*+}$ mesons. 
The experimental error was dominated by varying the cut on the scaled
energy of the D$^{*+}$ mesons which was required to reduce the background
from c quark events coming from cascade decays of b hadrons.
The measurement of \rasbasu\ achieved a 1.5\% precision, the error being
dominated by the theoretical systematic error. The largest theoretical errors
were due to uncertainties in the b quark mass and  
the renormalization scale factor $x_\mu$ .

In addition, we have presented a study of the effect of
heavy quark masses on global event shape variables. 
It was observed that the values of
\as\ determined from $y_{23}$, $1-T$, $B_W$ and $C$
were reduced by 5 to 7\%
when these event shapes were measured
with a sample of b quark events and a massless QCD calculation was used.
The shape variable $M_H$ was found to be insensitive to the b quark mass.

\section{Acknowledgements}
\par
We particularly wish to thank the SL Division for the efficient operation
of the LEP accelerator at all energies
 and for their continuing close cooperation with
our experimental group.  We thank our colleagues from CEA, DAPNIA/SPP,
CE-Saclay for their efforts over the years on the time-of-flight and trigger
systems which we continue to use.  In addition to the support staff at our own
institutions we are pleased to acknowledge the  \\
Department of Energy, USA, \\
National Science Foundation, USA, \\
Particle Physics and Astronomy Research Council, UK, \\
Natural Sciences and Engineering Research Council, Canada, \\
Israel Science Foundation, administered by the Israel
Academy of Science and Humanities, \\
Minerva Gesellschaft, \\
Benoziyo Center for High Energy Physics,\\
Japanese Ministry of Education, Science and Culture (the
Monbusho) and a grant under the Monbusho International
Science Research Program,\\
Japanese Society for the Promotion of Science (JSPS),\\
German Israeli Bi-national Science Foundation (GIF), \\
Bundesministerium f\"ur Bildung, Wissenschaft,
Forschung und Technologie, Germany, \\
National Research Council of Canada, \\
Research Corporation, USA,\\
Hungarian Foundation for Scientific Research, OTKA T-016660, 
T023793 and OTKA F-023259.\\

%
%
\clearpage

\section*{ Tables }


\begin{table}[ht]
\begin{center}

\begin{tabular}{|c||c|c|c|} \hline
          &  uds fraction  & charm fraction& bottom fraction\\ \hline \hline
 uds-tag  &$0.862\pm0.010$ &$0.119\pm0.003$&$0.019\pm0.011$ \\ \hline
 c-tag    &$0.246\pm0.005$ &$0.548\pm0.018$&$0.206\pm0.017$ \\ \hline
 b-tag    &$0.008\pm0.002$ &$0.028\pm0.004$&$0.964\pm0.019$ \\ \hline
\end{tabular} 

\end{center}
\caption{The fraction of uds, c and b quark events in each
flavour-tagged event sample. The errors are the combined statistical
and systematic errors.}
\label{tab:flavor}
\end{table}


\begin{table}[ht]
\begin{center}
\begin{tabular}{|c|c|c|c|c|}
\hline
cut   &  3 prong   & semileptonic  & satellite   & 5 prong \\
\hline
\hline
$x_{D^{*+}}$ & $0.4$--$1.0$ & $0.4$--$1.0$ & $0.4$--$1.0$ & $0.5$--$1.0$ \\
\hline
$M_{D^{0}}$ [GeV]& $1.79$--$1.94$ & $1.20$--$1.80$ & $1.41$--$1.77$ & $1.79$--$1.94$ \\
\hline
$\Delta M$ [GeV]& $0.142$--$0.149$ & $0.140$--$0.162$ & $0.141$--$ 0.151$ & $0.142$--$0.149$ \\
\hline
\end{tabular} 
\end{center}
\caption{
List of cuts used in the D$^{*+}$ reconstruction.
Note that both the scaled energy $x_{D^{*+}}$ and the mass difference
$\Delta M$ are effective quantities, calculated from the
reconstructed tracks only.
The exact meaning of the different quantities
is explained in the text.}
\label{tab:dstarcuts}
\end{table}


\begin{table}[t]
\begin{center}
\begin{tabular}{|c|c|c|c|c|c|} \hline
Event shape &Fit Range & \asu & \rascasu & \rasbasu & $\chi^2$/d.o.f.    \\ \hline\hline
 $y_{23}$ &0.015~--~0.185& $0.1263\pm0.0006$ & $1.017 \pm0.034 $ & $1.004 \pm0.007$& 1.4 \\ \hline
 $1-T$ &0.11~--~0.29 & $0.1451\pm0.0011$ & $0.967 \pm0.055 $ & $0.993 \pm0.011$& 2.2 \\ \hline
 $M_H$ &0.26~--~0.42 & $0.1344\pm0.0009$ & $1.004 \pm0.047 $ & $1.008 \pm0.009$& 2.5 \\ \hline
 $B_W$ &0.09~--~0.21 & $0.1313\pm0.0007$ & $1.006 \pm0.039 $ & $1.003 \pm0.008$& 4.4 \\ \hline
 $C$   &0.35~--~0.67 & $0.1424\pm0.0010$ & $0.986 \pm0.050 $ & $0.982 \pm0.011$& 1.6 \\ \hline
\end{tabular} 
\end{center}
\caption{Results of the $\chi^2$ fits
with the renormalization scale factor fixed: $x_\mu = 1$. The errors are 
statistical.}
\label{tab:oas2}
\end{table}

\begin{table}[t]
\begin{center}
\begin{tabular}{|c|c|c|c|c|c|} \hline
Event shape & \asu & \rascasu & \rasbasu & $x_\mu$& $\chi^2$/d.o.f.    \\ \hline\hline
 $y_{23}$ & $0.1188\pm0.0010$ & $1.008 \pm0.030 $ & $0.991 \pm0.006$ & $0.266 \pm 0.066$ & 1.0\\ \hline
 $1-T$ & $0.1163\pm0.0017$ & $0.963 \pm0.036 $ & $0.979 \pm0.007$ & $0.083 \pm 0.017$ & 0.5\\ \hline
 $M_H$ & $0.1195\pm0.0006$ & $0.990 \pm0.037 $ & $0.992 \pm0.008$ & $0.088 \pm 0.021$ & 1.2\\ \hline
 $B_W$ & $0.1191\pm0.0006$ & $0.995 \pm0.033 $ & $0.995 \pm0.007$ & $0.086 \pm 0.011$ & 1.8\\ \hline
 $C$   & $0.1125\pm0.0014$ & $0.973 \pm0.032 $ & $0.984 \pm0.010$ & $0.041 \pm 0.014$ & 1.1\\ \hline
\end{tabular} 
\end{center}
\caption{Results of the $\chi^2$ fits
with the renormalization scale factor $x_\mu$ as a free parameter. 
The errors are statistical. The fit ranges are given in Table~\ref{tab:oas2}. }
\label{tab:oas2free}
\end{table}


\begin{table}[ht]
\begin{center}
\begin{tabular}{|l||c|c|c|c|c|c|c|} \hline
Event shape        & $y_{23}$     & $1-T$     & $M_H/\sqrt{s}$     & $B_W$     & $C$       &Mean\\ \hline\hline 
$\alpha_s^c/\alpha_s^{uds}$ &{\bf 1.013}&{\bf 0.965}&{\bf 0.997}&{\bf 1.001}&{\bf 0.979}&{\bf 0.997}\\ \hline\hline
Statistical error  &$\pm 0.032$&$\pm 0.045$&$\pm 0.042$&$\pm 0.036$&$\pm 0.041$&$\pm 0.038$\\ \hline\hline
Flavour composition&$\pm 0.002$&$\pm 0.002$&$\pm 0.001$&$\pm 0.001$&$\pm 0.002$&$\pm 0.001$\\ \hline
 Clusters only     &$   +0.015$&$   +0.011$&$   +0.014$&$   +0.014$&$   +0.033$&$   +0.017$\\ \hline
Tracks only        &$   +0.009$&$   +0.012$&$   +0.037$&$   +0.006$&$\pm 0.001$&$   +0.012$\\ \hline
cos$\theta_{\rm Th} < 0.7$&$   +0.005$&$   -0.008$&$   +0.027$&$   +0.010$&$   +0.022$&$   +0.011$\\ \hline
$N_{ch} \ge 7$     &$  < 0.001$&$   +0.001$&$   +0.001$&$   +0.001$&$   +0.001$&$   +0.001$\\ \hline
Fit range          &$\pm 0.008$&$\pm 0.016$&$\pm 0.010$&$\pm 0.013$&$\pm 0.003$&$\pm 0.004$\\ \hline
$x_{D^*+}$ range    &$\pm 0.009$&$\pm 0.045$&$\pm 0.020$&$\pm 0.008$&$\pm 0.037$&$\pm 0.019$\\ \hline\hline
Experimental syst. &$\pm 0.022 $&$\pm 0.051 $&$\pm 0.053 $&$\pm 0.024 $&$\pm 0.055 $&$\pm 0.030$\\ \hline\hline
Lund frag. b       &$\pm 0.003$&$\pm 0.001$&$\pm 0.002$&$\pm 0.002$&$\pm 0.001$&$\pm 0.001$\\ \hline
Peterson frag. $\epsilon_c$&$\pm 0.010 $&$\pm 0.003 $&$\pm 0.002 $&$\pm 0.001 $&$\pm 0.003$&$\pm 0.003$\\ \hline
Peterson frag. $\epsilon_b$&$\pm 0.005 $&$\pm 0.004 $&$\pm 0.004 $&$\pm 0.004 $&$\pm 0.003$&$\pm 0.004$\\ \hline
$Q_0$              &$\pm 0.006 $&$\pm 0.001 $&$ <  0.001 $&$\pm 0.003 $&$\pm 0.003 $&$\pm 0.003$\\ \hline
$\sigma_q$         &$\pm 0.004 $&$\pm 0.003 $&$\pm 0.014 $&$ <  0.001 $&$\pm 0.005 $&$\pm 0.002$\\ \hline
Scale uncertainty  &$\pm 0.004 $&$\pm 0.002 $&$\pm 0.007 $&$\pm 0.005 $&$\pm 0.006 $&$\pm 0.007$\\ \hline
ARIADNE            &$  - 0.008 $&$ <  0.001 $&$  - 0.001 $&$  - 0.004 $&$  - 0.001 $&$  - 0.004$\\ \hline
c quark mass       &$\pm 0.008 $&$\pm 0.005 $&$ <  0.001 $&$\pm 0.005 $&$\pm 0.006 $&$\pm 0.005$\\ \hline
b quark mass       &$\pm 0.001 $&$ <  0.001 $&$ <  0.001 $&$\pm 0.001 $&$ <  0.001 $&$ <  0.001$\\ \hline\hline
Theoretical error  &$\pm 0.018 $&$\pm 0.008 $&$\pm 0.016 $&$\pm 0.010 $&$\pm 0.011 $&$\pm 0.012$\\ \hline\hline
Total error        &${\bf \pm 0.043 }$&${\bf \pm 0.069 }$&${\bf \pm 0.070 }$&${\bf \pm 0.044 }$&${\bf \pm 0.069 }$&${\bf \pm 0.050 }$\\ \hline
\end{tabular} \end{center}
\caption{Systematic errors on the ratio $\alpha_s^c/\alpha_s^{uds}$. 
Where a signed value is quoted, this indicates the
direction in which the ratio changed with respect to the default
analysis when a certain feature of the analysis was changed. 
See the text for a description of the systematic errors.}
\label{tab:sys_ascasu}
\end{table}


\begin{table}[ht]
\begin{center}
\begin{tabular}{|l||c|c|c|c|c|c|c|} \hline
Event shape                & $y_{23}$     & $1-T$     &$M_H/\sqrt{s}$ & $B_W$ & $C$       & Mean \\ \hline\hline 
$\alpha_s^b/\alpha_s^{uds}$&{\bf 0.997}&{\bf 0.986}&{\bf 1.000}&{\bf 0.999}&{\bf 0.983}&{\bf 0.993}\\ \hline\hline
Statistical error  &$\pm 0.007$&$\pm 0.009$&$\pm 0.008$&$\pm 0.008$&$\pm 0.010$&$\pm 0.008$\\ \hline\hline
Flavour composition&$\pm 0.001$&$\pm 0.001$&$ <  0.001$&$ <  0.001$&$\pm 0.001$&$ <  0.001$ \\ \hline
 Clusters only     &$   -0.004$&$   +0.001$&$   +0.001$&$   +0.001$&$\pm 0.001$&$ <  0.001$ \\ \hline
Tracks only        &$   -0.006$&$   +0.002$&$   +0.003$&$   +0.001$&$   -0.001$&$   -0.001$ \\ \hline
cos$\theta_{\rm Th} < 0.7$&$ <  0.001$&$   -0.001$&$   +0.002$&$   +0.001$&$   +0.002$&$   +0.001$ \\ \hline
$N_{ch} \ge 7$     &$ <  0.001$&$ <  0.001$&$ <  0.001$&$  < 0.001$&$  < 0.001$&$  < 0.001$ \\ \hline
Fit range          &$\pm 0.016$&$\pm 0.008$&$\pm 0.009$&$\pm 0.003$&$\pm 0.003$&$\pm 0.005$ \\ \hline
$x_{D^*+}$ range    &$\pm 0.001$&$\pm 0.004$&$\pm 0.002$&$\pm 0.002$&$\pm 0.003$&$\pm 0.002$ \\ \hline\hline
Experimental syst. &$\pm 0.017$&$\pm 0.010$&$\pm 0.011$&$\pm 0.004$&$\pm 0.005$&$\pm 0.006$ \\ \hline\hline
Lund frag. b       &$\pm 0.001$&$\pm 0.005$&$\pm 0.003$&$\pm 0.001$&$\pm 0.005$&$\pm 0.001$ \\ \hline
Peterson frag. $\epsilon_c$ &$\pm 0.003$&$\pm 0.007$&$\pm 0.005$&$\pm 0.002$&$\pm 0.001$&$\pm 0.003$ \\ \hline
Peterson frag. $\epsilon_b$ &$\pm 0.004$&$\pm 0.004$&$\pm 0.003$&$\pm 0.001$&$\pm 0.004$&$\pm 0.002$ \\ \hline
$Q_0$              &$\pm 0.004$&$\pm 0.008$&$\pm 0.003$&$\pm 0.003$&$\pm 0.005$&$\pm 0.003$ \\ \hline
$\sigma_q$         &$\pm 0.003$&$\pm 0.009$&$\pm 0.005$&$ <  0.001$&$\pm 0.001$&$\pm 0.001$ \\ \hline
Scale uncertainty  &$\pm 0.006$&$\pm 0.007$&$\pm 0.008$&$\pm 0.004$&$\pm 0.001$&$\pm 0.005$ \\ \hline
ARIADNE            &$  + 0.004$&$  - 0.001$&$  - 0.001$&$  + 0.008$&$  - 0.002$&$  + 0.002$\\ \hline
c quark mass       &$ <  0.001$&$ <  0.001$&$ <  0.001$&$ <  0.001$&$ <  0.001$&$ <  0.001$ \\ \hline
b quark mass       &$\pm 0.012$&$\pm 0.008$&$\pm 0.005$&$\pm 0.010$&$\pm 0.010$&$\pm 0.008$ \\ \hline\hline
Theoretical error  &$\pm 0.016$&$\pm 0.018$&$\pm 0.013$&$\pm 0.014$&$\pm 0.013$&$\pm 0.011$ \\ \hline\hline
Total error        &${\bf \pm 0.025}$&${\bf \pm 0.022}$&${\bf \pm 0.019}$&${\bf \pm 0.016}$&${\bf \pm 0.017}$&${\bf \pm 0.014}$ \\ \hline
\end{tabular} \end{center}
\caption{Systematic errors on the ratio $\alpha_s^b/\alpha_s^{uds}$. 
Where a signed value is quoted, this indicates the
direction in which the ratio changed with respect to the default
analysis when a certain feature of the analysis was changed.
See the text for a description of the systematic errors.}
\label{tab:sys_asbasu}
\end{table}

\clearpage
\section*{ Figures }

\end{document}